\documentclass[manuscript]{aastex}

\slugcomment{To appear is PASJ}

\begin{document}

\title{How Evolved are the Mass Donor Stars in
Cataclysmic Variables?}

\author{Steve B. Howell}
\affil{Astrophysics Group, Planetary Science Institute, Tucson, AZ  85705}
\email{howell@psi.edu}


\maketitle

\begin{abstract}
Recent spectroscopic observations
have identified several cataclysmic variables
non-solar metal abundances.
We present theoretical models which
examine the level of core evolution expected for CV secondaries
prior to contact.
Our results indicate that few
secondary stars evolve past 10\%
of their main sequence lifetime prior to the initiation
of mass transfer; a result which is in agreement with present day observations.
Thus, the non-solar metal abundances
observed must
be due to nuclear burning by-products
accreted by the secondary star during common envelope sweeping
prior to initial contact or
ejecta collected during classical nova outbursts.
\end{abstract}

\section{Introduction}

Cataclysmic variables (CVs) are semi-detached binaries consisting of 
a white dwarf (WD) primary, a low
mass (assumed) main sequence mass donor secondary star (M$_{don}$), and accretion 
phenomena (stream, disk, etc.).
Recent spectroscopic evidence has shown that non-solar metal 
abundances are present in some
CVs. Cheng et al. (1997) 
determined that the carbon abundance was 5 times solar, 
nitrogen was 3 times solar, and silicon was $<$0.1 solar for the white dwarf in WZ
Sge. These authors concluded
that the most likely source of
this non-solar composition 
material was the mass transfered from the 
secondary star. Harrison et al. (2000)
present IR spectra of three long period systems (U Gem, SS Aur, SS Cyg) and
note the apparent weakness of the CO bandheads possible implying an under abundance of
carbon. Long and Gilliland (1999) analyzed the WD spectrum of U Gem and determined
that nitrogen was 4 times solar while C was $<$0.1 solar.
The polar BY Cam has been observed in the FUV spectral region and 
that analysis (Mouchet et al., 2001) has revealed increased nitrogen and
decreased carbon abundances 
in agreement with CNO processing in thermonuclear runaways. Finally,
Howell and Ciardi (2001) observed two ultra-short period CVs (LL And and EF Eri)
and their IR spectra show very weak CO absorption, indicative of a 
less than solar carbon abundance.

If we assume that the component stars in a CV 
are born with either solar composition or if
very old, low metal abundance,
some mechanism is required to produce the observed
anomalies.
Nuclear core evolution within the secondary star could possibly provide this 
mechanism 
if He burning occurs, that is, if the secondary star reaches
an advanced main sequence state or post-main sequence phases in its evolution. 
Prior to this time, only He enrichment
can occur as the core nuclear burning in these low mass stars is via the
pp-cycle which does not produce any abundance changes among the CNO elements. 

Marks and Sarna (1998) performed a detailed theoretical study of the possible effects
on the surface abundances of the secondary star due to sweeping up common envelope
(CE) material or accreting classical nova ejecta; both cases would place 
thermonuclear processed material onto the secondary star. 
The work by Marks \& Sarna only
considered initially massive secondary stars (1.0-1.5 M$_{\odot}$)
whereas Howell, Nelson, \& Rappaport (2001; hereafter HNR) have 
shown that massive secondaries are likely to represent only a small fraction of
all CVs (See \S2.2). 
Recurrent novae, super-soft X-ray sources, and very long period
CVs (P$_{orb}$ $>$7 hr) are examples of systems with massive and/or evolved
secondary stars.

In order to test the possibility that nuclear core
evolution might play a role in the chemical anomalies observed,
we have
undertaken a new 
population synthesis calculation of over 3 million primordial
binaries, of which $\sim$20,000 end up as mass
transferring CVs.
We then statistically examine 
the age of the mass donors prior
to contact to determine what level of core evolution has occured. 

\section{Secondary Star Evolution}

CVs begin as deteched binaries containing two main sequence 
stars which then pass through a
common envelope (CE) stage during which the lower mass (secondary) star 
orbits within
the extended outer atmosphere of the faster evolved, more massive star. 
The core of the more massive or primary 
star will become the white dwarf that will accompany the
low mass star throughout the remainder of its life.
CVs which initially contain two low mass components generally require 
a long time to
come into contact after the CE phase. In such systems, 
the secondary star's main sequence lifetime is longer than
a Hubble time so no significant main sequence core evolution is expected.

If a binary reaches the semi-detached state, ie., mass transfer occurs, at a long
orbital period, the masses of the component stars both must be 
fairly high, $>$ 0.4 M$_{\odot}$ (see HNR). These long period CV's are the systems
for which significant core evolution {\it could} occur in the mass donor
prior to contact and within the lifetime of
the Galaxy (10 GY). Baraffe \& Kolb (2000) examined evolutionary models 
and showed that for
CVs with orbital periods longer than about 7 hr, 
the secondary stars have initial masses of $>$1M$_{\odot}$
and they are 
evolved prior to contact.
These authors used an empirical relation between spectral type and I-K color 
(derived from
stellar models) to obtain a spectral type-orbital period relation for a given
mass-effective temperature-luminosity model of the secondary star. 

The relation between spectral type and secondary star mass (or P$_{orb}$--M$_2$,
P$_{orb}$--Sp. Type)
is fairly accurate for CVs with orbital periods below the
period gap and for those with P$_{orb}>$6 hr (See Warner 1995; HNR). 
However, for those CVs with periods
in the range of 3-6 hr, the secondary star is bloated due to its 
non-thermal equilibrium state making it
larger in radius when compared with a main sequence
star of the same mass. The effect is illustrated in Figure 1 through the use of 
the main
sequence (solid line) and proper mass donor (dashed line) P$_{orb}$--M$_2$
relationships. Four example secondary stars, in pairs at equal orbital period,
are shown in Figure 1. Table 1 provides some theoretical intrinsic 
properties for the stars in each pair
and shows that main sequence values are unreliable to use within the 3-6 hr
period range. For example, a CV with P$_{orb}$=3.2 hr would be estimated to have
a mass of 0.37 M$_{\odot}$ while its most likely 
true mass is only 0.25 M$_{\odot}$.
In addition, using an assumed spectral type for the secondary star 
based solely on 
P$_{orb}$ will also be in error.
Thus, the effect of bloating on the stars within the orbital period
range of 3-6 hr makes the use of
orbital period alone invalid as a pointer to the intrinsic properties of the
mass donors.

\subsection{Secondary Star Evolution After Contact}

The evolution for long period CVs (P$_{orb}>$3 hr)
and the corresponding rapid mass loss of the
secondary star, as the orbital period decreases,
are illustrated for three example systems in Figure 2 and Table 2. 
It can be seen that even an initially 
massive secondary star ($\sim$ 1M$_{\odot}$) 
quickly (within 260 million years) becomes a low mass star (0.25M$_{\odot}$).
Most secondary stars within long period CVs lose 70\% or more 
of their initial mass
within 0.4 GY.
Even a casual examination of these well known timescales shows
that the mass trasnfer rate above the period gap is so large that the secondary
stars have no chance of nuclear core evolution during their contact phase.
Details of the evolution process can be found in HNR.

Thus, if nuclear core evolution within the secondary star
does contribute significantly to the non-solar
metal abundances observed, 
it must happen prior to contact (ie., before the start of mass transfer)
during the time period in which the secondary 
evolves essentially as a single main sequence star. 

\subsection{Secondary Star Evolution Prior to Contact}

Our population synthesis model starts by chosing
a sample of primodial binaries with 0.8 M$_{\odot}$ $<$ M$_1$ $<$ 
8 M$_{\odot}$ and then picking appropriate secondary stars for each based on
observational work and 
stability conditions detailed in HNR. Since the primary star
evolves into a white dwarf with a mass that can be as large as 
1.4 M$_{\odot}$, the allowed
secondaries can have initial masses of up to $\sim$1.8 M$_{\odot}$. 
While HNR's binary
evolution code does not deal with secondaries of initial mass greater than 1
M$_{\odot}$, an examination of the number of possible systems ignored
due to this constraint 
reveals that it represents only a small fraction of the present day CV population.

Using Monte Carlo techniques, ten new samples of 200,000 initial primordial 
binaries each 
were generated for this work. The number of produced binaries which
contained secondary stars with initial masses of 
M$_2$$>$ 1 M$_{\odot}$ was found to be near 0.5\% in each set.
Thus, of the 3 million primordial binaries used by HNR 
in their work, the 0.5\% value amounts to having ignored $\sim$ 22,000
systems which started with massive secondaries. 
Unless there is some as
yet unknown reason why primordial binaries with M$_2$$>$ 1 M$_{\odot}$ have a
greater efficiency for producing mass transferring CVs than their lower mass
counterparts, these 22,000 binaries would produce for only 45 present day mass
transferring CVs, if they had been properly evolved through their lifetime.
This number represents $<<$ 1\% of the total HNR sample, 
all of which would have started mass transfer 
at P$_{orb}>$6 hr.

How does this theoretical result compare with observation? 
The Ritter \& Kolb (1998) catalogue
lists 318 CVs which have a known orbital period. Of these, 10\% have P$_{orb}>$6
hr and probably contain secondary stars of mass greater than 1 M$_{\odot}$.
Ten of these systems are novae, likely indicating a massive white dwarf which is
needed to successfully pair with a massive secondary. Of the six long period
systems for which a secondary star spectral type is given, only one is a main
sequence star, the rest are post-main sequence objects. Thus, we find strong
evidence that very long period CVs contain evolved secondaries of mass $>$ 
1 M$_{\odot}$, a result in agreement with Baraffe \& Kolb (2000). Of the 
$\sim$1000 CVs known at present, those with long orbital periods are 
brighter and highly favored
for observational discovery. Taken at face value, one would 
infer that the 30 or so CVs known in Ritter \& Kolb with massive
secondaries represent 3\% of the total present day population of 1000. 
However, 
the extreme observational bias present for discovery of long period, bright CVs
compared with the vast majority of fainter, shorter period systems. 
Figure 9 in HNR
shows that CVs with P$_{orb}>$ 6 hr are $\sim$1000 times more likely to 
be discovered
than those with P$_{orb}<$ 6 hr, making the observed 3\% actually represent
$<<$ 1\% of the
true population. This small predicted value is in quite good 
agreement with the theoretical result 
given above based on our population systhesis selection of primodial binaries 
and their
subsequent evolution.
Given that the vast majority of CVs start their lives with
orbital periods of $<$6 hr and secondary stars with initial masses of
$<$ 1M$_{\odot}$, we now examine the likelihood of nuclear core evolution
in the secondary stars of these, the majority of CVs.

We present in Figure 3 the percent of the main sequence lifetime 
reached prior to contact 
for each secondary star in our model population for those binaries which 
were successful at becoming
mass transferring CVs within the lifetime of the Galaxy. 
Figure 3 contains about 20,000
systems and is based on the theoretical work presented in HNR.
We see that most secondary stars come into contact 
early in their main sequence life, generally within about 5-10\% of their ZAMS
time.
Less than
5\% of all donors evolve beyond one-fifth of their main sequence 
lifetime prior to Roche Lobe
contact. 
The peak in the distribution seen near 0.7 $M_{\odot}$ is due to the trade off
between earlier contact time for more massive (larger) secondary stars 
(these must be
paired with a massive WD), and the much longer main sequence
lifetimes for secondaries 
initially
of mass $<$ 0.7 $M_{\odot}$. A CV which comes into contact
with a 0.7 $M_{\odot}$ secondary typically does so at an orbital period near
5.0-5.5 hr.
Figure 3 illustrates that even massive
secondaries are unlikely to advance past 20\% of their main sequence lifetime
prior to contact.

\section{Conclusion}

From the results presented in \S2.2
it is unlikely that nuclear core evolution (ie., He burning) 
{\it prior} to contact
can be an important component leading to non-solar metal abundances within the
atmospheres of the vast majority of present day secondary stars in CVs.
Additionally, the 
rapidity of the secondary star evolution to very low mass after contact
can produce no
core evolution of any kind beyond that already determined for the star prior to
contact. 
Thus, we are forced to conclude
that no significant nuclear core evolution, specifically CNO abundance changes,
will occur for the vast majority
of the 
secondary stars in the present day population of CVs. 
Therefore, any observed atmospheric abundance anomolies
must be attributed to either material swept up by the secondary star during CE
evolution or material accreted by the secondary star during classical novae
eruptions on the white dwarf. Primordial abundances could 
also be considered, however, even for CVs of advanced age
this seems to only allow for lower metal abundances not
enhanced values as have been observed. 

Marks and Sarna (1998) and Sarna et al. (1995)
have modeled the CNO
burning processes of both CE and nova thermonuclear reactions. 
They find 
that if the secondary star accretes $\sim$0.05 $M_{\odot}$ of material
during the CE phase, its $^{12}$C/$^{13}$C and $^{16}$O/$^{17}$O ratios would 
fall to
near half the solar value. 
Livio \& Truran (1994)
and Kovetz \& Prialnik (1997) also modeled turbulent nuclear burning and 
CNO redistribution via nova
processes. All of these authors generally agree that an
enhancement of nitrogen, stationary or decreased oxygen, 
and depletion of carbon are 
the by-products of the CNO nuclear processes made available to the secondary star 
during
the CE stage or following the dredge-up of matter from an underlying CO core during
a classical nova outburst. 
The non-solar CNO abundances observed in CVs appear to agree with the trends
outlined in these
theoretical calculations.

Determinations of accurate CV metallicities 
will require high S/N (50-100) continuum spectroscopic observations.
For example, even at 
2.2 microns the secondary star is often not the dominate
flux source. Any spectroscopic observations aimed at abundance determinations must
realistically account for rapid rotation effects 
and accretion flux contributions, both of which distort spectral lines,
in order to properly measure absoprtion features.
Given that low mass, He core white dwarfs are unlikely to produce
nova outbursts, spectroscopic identification of non-solar 
abundances (consistant with nuclear
by-products) in such a CV would provide strong 
evidence in favor of significant pollution of the secondary star composition
during the CE phase.

David Ciardi is thanked for reading and commenting on an early draft 
of this work.
SBH acknowledges partial support for this research from NSF grant 
AST 98-19770 and NASA Theory grant NAG5-8500. 

\section{References}

Baraffe, I., \& Kolb, U., 2000, MNRAS, 318, 354.

Beuermann, K., Baraffe, I., Kolb, U., \& Weichhold, M., 1998, A\&A, 339, 518.

Cheng, F., Sion, E., Szkody, P., \& Haung, M., 1997, ApJ, 484, L149.

Cox, A., 2000, ``Astrophysical Quantities", AIP press, Chps. 7 \& 15.

Harrison, T., McNamara, B., Szkody, P., \& Gillland, R., 2000, AJ, 120, 2649.

Howell, S. B., Nelson, L., \& Rappaport, S, 2001, ApJ, 550, 897. [HNR]

Howell, S. B., \& Ciardi, D., 2001, ApJL, 550, L57.

Kovetz, A., \& Prialnik, D., 1997, ApJ, 477, 356.

Livio, M., \& Truran, J., 1994, ApJ, 425, 797.

Long, K., \& Gilliland, R., 1999, ApJ, 511, 916.

Marks, P., \& Sarna, M., 1998, MNRAS, 301, 699. 

Miller, G., \& Scalo, J., 1979, ApJS, 41, 513.

Mouchet, M., Bonnet-Bidaud, J.-M., Beuermann, K., De Martino, D., Ferlet, R.,\\
Fried,
R., Howell, S. B., Lecaveleir Des Etangs, A., Mukai, K., Porquet, E., Roueff, E., \\
\&
Szkody, P., 2001, ``Cosmic Evolution", World Scientific, in press.

Ritter, H., \& Kolb, U., 1998, A\&AS, 129, 83.

Sarna, M., Dhillon, V., Marsh, T., \& Marks, P., 1995, MNRAS, 272, L41.

Smith, D., \& Dhillon, V., 1998, MNRAS, 301, 767.

\begin{deluxetable}{ccccc}
\tablewidth{4.5in}
\tablenum{1}
\tablecaption{Model Stars in Figure 1\tablenotemark{a}}
\tablehead{
\colhead{Parameter}
  &\colhead{Model A}
  &\colhead{Model B}
  &\colhead{Model C}
  &\colhead{Model D}
}
\startdata
Mass (M$_{\odot}$) & 0.37 & 0.25 & 0.47 & 0.37 \\
Radius (M$_{\odot}$) & 0.38 & 0.32 & 0.46 & 0.42 \\
P$_{orb}$ (hr) & 3.2 & 3.2 & 4.0 & 4.0 \\
log g & 4.84 & 4.82 & 4.78 & 4.76 \\
T$_{eff}$ (K) & 3300 & 3600 & 3700 & 3800 \\
Sp. Type(1)\tablenotemark{b} & M5.2 & M5.2 & M3.2 & M3.2 \\
Sp. Type(2)\tablenotemark{c} & M3 & M2 & M0.5 & M0 \\
B-V(1)\tablenotemark{d} & 1.62 & 1.62 & 1.55 & 1.55 \\
B-V(2)\tablenotemark{d} & 1.58 & 1.49 & 1.45 & 1.4 \\
I-K(1)\tablenotemark{e} & 2.5 & 2.5 & 2.3 & 2.3 \\
I-K(2)\tablenotemark{e} & 2.2 & 2.0 & 1.71 & 1.65 \\
\enddata
\tablenotetext{a}{Models are calculated based on the work presented in HNR.}
\tablenotetext{b}{From Smith \& Dhillon (1998) relation of spectral type vs.
P$_{orb}$.}
\tablenotetext{c}{Based on MK Spectral Classification assuming main 
sequence stars
with the given T$_{eff}$ (Cox 2000).}
\tablenotetext{d}{Based on T$_{eff}$ and spectral type for main sequence stars
(Cox
2000).}
\tablenotetext{e}{Based on spectral type, from Beuermann et al., (1998).}
\end{deluxetable}

\begin{deluxetable}{cccc}
\tablewidth{4.5in}
\tablenum{2}
\tablecaption{Time (GY) Since Initial Contact 
for the Three Models Shown in Figure 2}
\tablehead{
\colhead{Number}
  &\colhead{M$_1$=1.4M$_{\odot}$}
  &\colhead{M$_1$=0.6M$_{\odot}$}
  &\colhead{M$_1$=0.3M$_{\odot}$} \\
\colhead{in Fig. 2}
  &\colhead{M$_{don}$=1.0M$_{\odot}$}
  &\colhead{M$_{don}$=0.5M$_{\odot}$}
  &\colhead{M$_{don}$=0.2M$_{\odot}$} \\
}
\startdata
1 & 0.01 & 0.02 & 0.6 \\
2 & 0.07 & 0.09 & 2.3 \\
3 & 0.13 & 1.8 & 3.9 \\
4 & 0.27 & 4.2 & 5.5 \\
5 & 0.90 & 5.5 & -- \\ 
6 & 1.7 &  9.7 & -- \\
7 & 2.8 & -- & -- \\
8 & 3.5 & -- & -- \\
9 & 4.6 & -- & -- \\
\enddata
\end{deluxetable}


\begin{figure}
\plotone{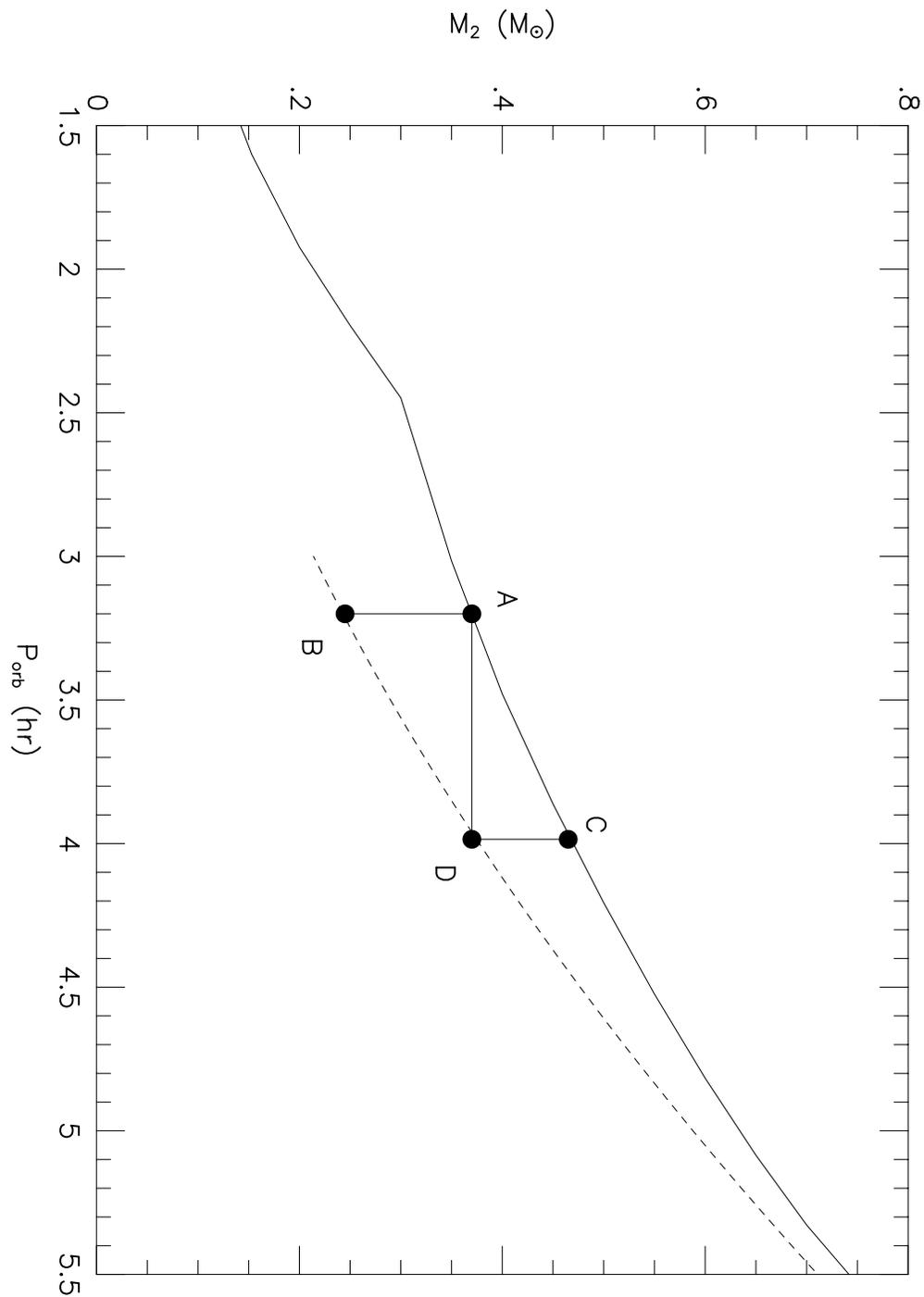}
\caption[]{An illustration of the effect of bloating on the secondary stars in
CVs. The solid line is the main sequence relation and the dashed line is that for
mass transferring secondaries in CVs. Below the period gap and above 6 hr, 
the two relations are
essentially the same. 
The four model secondaries have their parameters listed in Table 2. Bloating
is most extreme between orbital periods of 3-5 hr.}
\end{figure}

\begin{figure}
\plotone{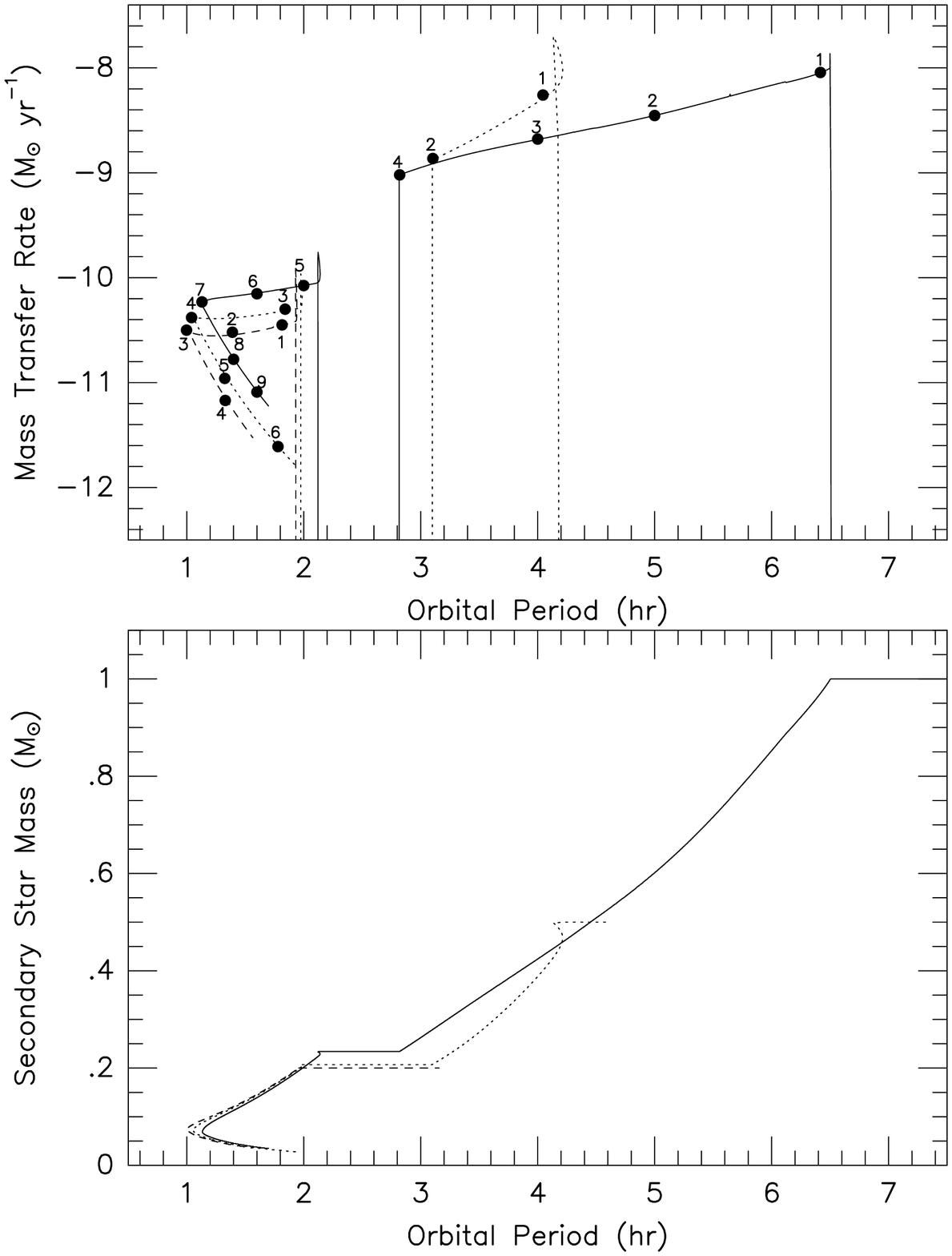}
\caption[]{The top panel presents three representative CV evolutions
in the orbital period, mass transfer plane. The three models are
(solid) a CV with M$_1$=1.4M$_{\odot}$ and (initially) M$_{don}$=1.0M$_{\odot}$, 
(dotted)
a CV with M$_1$=0.6M$_{\odot}$ and M$_{don}$=0.5M$_{\odot}$, and (dashed) a CV with
M$_1$=0.3M$_{\odot}$ and M$_{don}$=0.2M$_{\odot}$. The numbers refer to a time
history for each model as
listed in Table 2. The bottom panel shows the
same three models but in the orbital period, secondary star mass plane. 
The low mass binary does not
start mass transfer until it reaches an orbital period of less than 2 hr.
Note how
rapidly the the mass of the secondary star decreases for the long period CVs.}
\end{figure}

\begin{figure}
\plotone{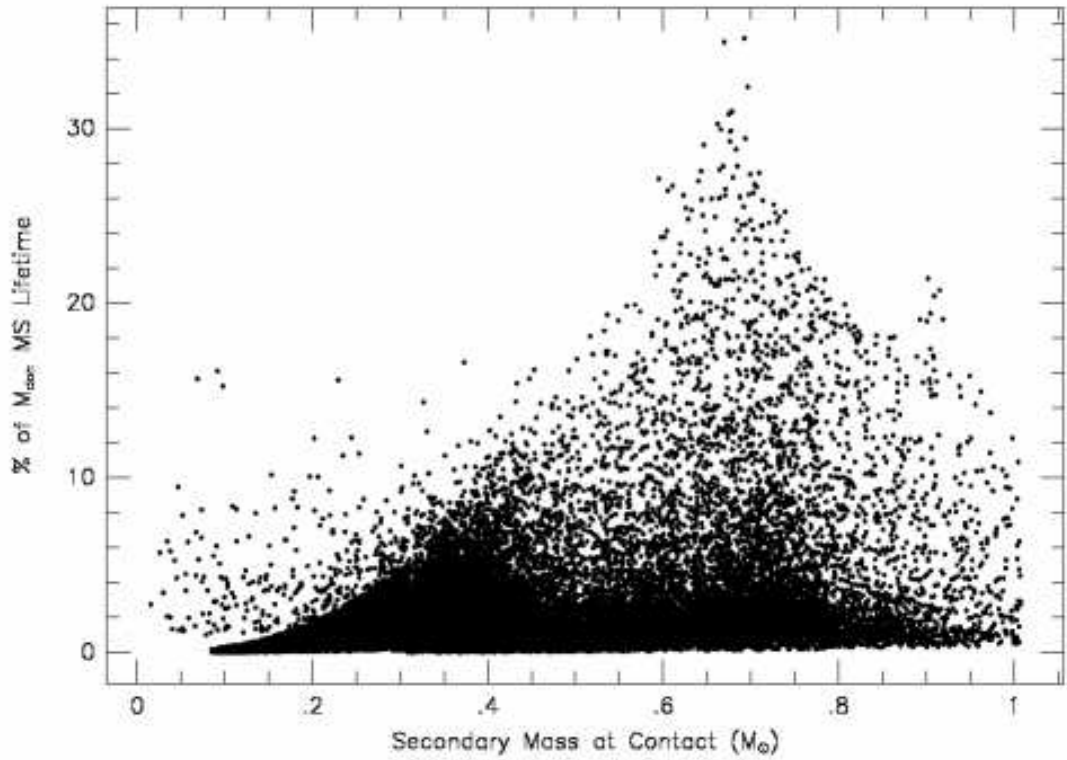}
\caption[]{The percent of the secondary star 
main sequence lifetime
which has passed prior to the beginning of mass transfer vs. the secondary star
mass at the time of initial contact. Most secondary stars
have initial Roche lobe contact early in their main 
sequence lives. 
See text for details.}
\end{figure}
\end{document}